\documentclass[twocolumn,aps,amsmath,amssymb,showpacs,pre]{revtex4-1}
\usepackage{graphicx}
\usepackage{amssymb}
\begin{document}
\title{Transfer-matrix study of a hard-square lattice gas with two 
kinds of particles and density anomaly} 
\author{Tiago J. Oliveira}
\email{tiago@ufv.br}
\affiliation{Departamento de F\'isica, Universidade Federal de Vi\c cosa, 36570-900, Vi\c cosa, MG, Brazil}
\author{J\"urgen F. Stilck}
\email{jstilck@if.uff.br}
\affiliation{Instituto de F\'{\i}sica and National Institute of Science and Technology for Complex Systems, Universidade Federal Fluminense, Av. Litor\^anea s/n, 24210-346, Niter\'oi, RJ, Brazil}   

\date{\today}

\begin{abstract}
Using transfer matrix and finite-size scaling methods, we study the thermodynamic behavior of a lattice gas with two kinds of particles on the square lattice. Only excluded volume interactions are considered, so that the model is athermal. Large particles exclude the site they occupy and its four first neighbors, while small particles exclude only their site. Two thermodynamic phases are found: a disordered phase where large particles occupy both sublattices with the same probability and an ordered phase where one of the two sublattices is preferentially occupied by them. The transition between these phases is continuous at small concentrations of the small particles and discontinuous at larger concentrations, both transitions are separated by a tricritical point. Estimates of the central charge suggest that the critical line is in the Ising universality class, while the tricritical point has tricritical Ising (Blume-Emery-Griffiths) exponents. The isobaric curves of the total density as functions of the fugacity of small or large particles display a minimum in the disordered phase. 
\end{abstract}


\maketitle

\section{Introduction}
\label{intro}
Models for fluids with only repulsive interactions have been studied in the literature for quite a long time \cite{hd86}, both on a lattice and in the continuum. Although attractive interactions are essential to produce liquid-gas transitions, models with only repulsive interactions may show transitions which resemble the melting of a solid phase. If the repulsive interactions are of the hard-core excluded volume type, all allowed configurations of the system have the same energy and thus the model is athermal. The continuous versions of these models are known as hard sphere models, and a fluid-solid transition was found in their phase diagram. It is worth recalling that in the seminal work by Metropolis et al., where Monte Carlo simulations were introduced, the hard sphere gas was studied \cite{hd86,mrrtt53}.

Athermal lattice models with excluded volume interactions have also been widely studied before. In these models the localization of particles is constrained to sites of a lattice. A model where a particle excludes others from only its site corresponds to an Ising lattice gas without the interaction term, and no phase transition is found. If a particle placed on a site of the square lattice excludes other particles from its four first neighbor sites, at low densities the sites of both sublattices are equally occupied, but as the density is increased one of the sublattices will be preferentially occupied by the particles. One may associate the disordered low density phase to a fluid and the ordered phase to a solid, so that the transition may be seen as a melting of a solid phase. This model has been thoroughly investigated by a variety of techniques \cite{b60}. On the square lattice, the universality class of the transition was object of some discussion in the literature \cite{b80}. Accurate estimates of the thermodynamic behavior of the model using transfer matrix and finite-size scaling techniques show a continuous transition between the disordered and the ordered phase, which is in the Ising universality class \cite{gb02}, as one should expect from symmetry considerations. This model may be generalized by increasing the range of the excluded volume interactions, and discontinuous phase transitions are found if this range is large enough. We refer to recent simulational investigations of this family of models where comprehensive surveys of the literature may be found \cite{fal07,rajesh14}. Although the transfer matrix formalism may lead to precise estimates for the critical behavior of such athermal models, Monte Carlo methods with cluster algorithms may also furnish good results. As examples, we mention the Ising lattice gas with first neighbor exclusion on the cubic \cite{hb96} and triangular \cite{zd08} lattices.

Here we consider another generalization of the lattice gas with first neighbor exclusion, introducing also small particles which exclude only the site they occupy. This mixed lattice gas was studied, using series expansion techniques on the square lattice, by Poland \cite{p84}. He found evidences for a tricritical point in the phase diagram of the model: while for low densities of small particles the transition is continuous, it becomes discontinuous as the density is increased. A slight modification of this model, where the large particles occupy elementary squares of the lattice and the small particles are located on the center of edges, has been exactly solved in the grand canonical formalism. In a particular case of this model,  when the fugacities of the small ($z_1$) and large ($z_2$) particles obey the relation $z_2=(1+z_1)^2$, Frenkel and Louis were able to show that it may be mapped on the Ising model with vanishing magnetic field and thus its solution in two dimensions is known \cite{fl92}. It should be mentioned that the fact that a model very similar to the original can be mapped on the Ising model at a particular point of the critical line is a strong additional evidence that the transition is in the Ising universality class. In particular, for $z_1=0$ the Frenkel-Louis model corresponds to the model with only large particles when their fugacity is unitary. A variant of this model was proposed and studied by Lin and Taylor \cite{lt94}. In this model, the small particles are triangles such that one of their sides occupies an edge of the lattice and the opposite vertex is located at the center of an elementary square, so that up to four small particles may be placed in an elementary square of the lattice. If finite interaction energies between the large (square) particles on first-neighbor sites and between large (square) and small (triangle) particles which share a lattice edge are introduced, this model shows a lower critical solution point. We also notice that a decorated lattice model for hydrogen-bonded mixtures which can be mapped on the three-dimensional Ising model and which shows both upper and lower critical solution points has been studied by Wheeler and Anderson \cite{wa80}.

More recently, the solution in the grand-canonical formalism of the model on a Bethe lattice with arbitrary coordination number was obtained \cite{tj11}. The general features of the phase diagram are consistent with the findings of Poland \cite{p84}. In the parameter space defined by the two activities $z_2$ and $z_1$ of large and small particles, respectively, a continuous transition is found for small values of $z_1$, and it becomes discontinuous if $z_1$ is sufficiently large. Since the slope of the critical line is negative at low values of $z_1$, becoming positive at higher values, a re-entrant behavior of the transition is seen in this region of the phase diagram. At even higher values of the activity of small particles, the transition becomes discontinuous and thus a tricritical point is found. Another interesting feature of the model is that the isobaric curves, where the total density of particles is considered as a function of the density or the activity of small particles, shows a minimum in the fluid phase. We recall that perhaps the most studied situation in nature where a density anomaly is found happens in water close to the freezing point, where a maximum of the isobaric curves of the  density as a function of the temperature is seen. The Bethe lattice solution of the mixed lattice gas model leads to a minimum of the density as a function of another field-like thermodynamic variable: the activity of small particles. Although the physical situation in water is of course quite distinct from the model we study here, it is noteworthy that several recent studies of simple effective models for water suggest that the density anomaly may be due to effective interparticle interaction potentials with two length scales \cite{s10}, a feature which is also present in the mixed lattice gas model.

Through transfer matrix and finite-size scaling calculations, we show here that the thermodynamic behavior of the model on the square lattice is qualitatively the same found in the Bethe lattice solution. Our results suggest that the critical line is in Ising universality class, similarly to the model with large particles only, and that the tricritical point belongs to the tricritical Ising (Blume-Emery-Griffiths - BEG) class.

In Sec. \ref{defMT}, we define the model more precisely and describe its transfer matrix solution on strips of finite widths, for both periodic and helical boundary conditions. We also discuss the factorization of this transfer matrix in a product of sparse matrices, which reduces the computational effort to handle them. Our results for the phase transitions and other thermodynamic properties of the model in the two-dimensional limit may be found in Sec. \ref{resperiodic}. Section \ref{conc} is devoted to final discussions and conclusion. 

\section{Definition of the mixed lattice gas model and its transfer matrix}
\label{defMT}

\begin{figure}[!t]
\begin{center}
\includegraphics[width=8.5cm]{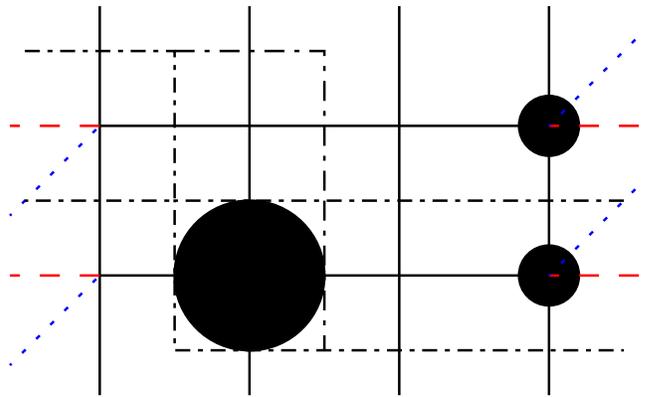}
\caption{(color on line) Strip of width $L=4$, with periodic (dashed-red) and helical (dotted-blue) boundary conditions. The lower and upper rows of sites determine the states of the transfer matrix. They are $(0,2,0,1)$ and $(0,0,0,1)$, respectively. The corresponding element is $z_2^{1/2}\,z_1$. The partially overlapping dot-dashed boxes are the states ($(2,0,1,0)$ and $(0,1,0,0)$)for helical boundary conditions, and the corresponding matrix element is $1$, since a single empty site is added.}  
\label{bc}
\end{center}
\end{figure}

We will use the transfer matrix formalism to study the model with both particles placed on a square lattice. This will be accomplished by solving the model on strips of finite width $L$. For periodic boundary conditions, the states of the transfer matrix will be defined by the configuration of the $L$ lattice sites in the same row of a cylinder. As an example, for $L=4$ we have a total of 26 states, but this number reduces to 9 if the rotation symmetry is considered. These states, with their multiplicity indicated between curly brackets,  are: $(0,0,0,0)-\left\lbrace 1 \right\rbrace$, $(0,0,0,1)-\left\lbrace 4 \right\rbrace$, $(0,0,0,2)-\left\lbrace 4 \right\rbrace$, $(0,0,1,1)-\left\lbrace 4 \right\rbrace$, $(0,1,0,1)-\left\lbrace 2 \right\rbrace$, $(0,1,0,2)-\left\lbrace 4 \right\rbrace$, $(0,1,1,1)-\left\lbrace 4 \right\rbrace$, $(0,2,0,2)-\left\lbrace 2 \right\rbrace$, and $(1,1,1,1)-\left\lbrace 1 \right\rbrace$. In our notation, $0$ represents an empty site, while $1$ and $2$ correspond to sites occupied by small and large particles, respectively. To avoid frustration in the solid phase, we restrict ourselves to even widths.  The transfer matrix is obtained considering two adjacent rows of $L$ sites in a particular configuration of particles and checking if the excluded volume interactions are satisfied. If this is the case, the corresponding element of the transfer matrix will be given by $T_L(i,j) = K z_1^{n_1/2}z_2^{n_2/2}$, where $n_1$ and $n_2$ are the total numbers of small and large particles in both rows, respectively, and $K$ is the number of ways (considering the excluded volume) of placing the row $j$ over the row $i$, by keeping $i$ fixed and rotating $j$. Defining the matrix elements in this way assures that the transfer matrix is hermitian. For $L=4$, some elements of the transfer matrix, with the states ordered as presented above, are given by:
\begin{eqnarray}
 T_4(1,2)=4 z_1^{1/2}, \quad & & T_4(2,2)= 4 z_1,  \\ \nonumber
 T_4(3,9)=0, \quad \quad \quad & &  T_4(5,8)=z_1 z_2.
\end{eqnarray}

Besides adopting periodic boundary conditions, we also considered helical boundary conditions, where the rightmost site of a row is linked to the leftmost site of the row above. Both boundary conditions are illustrated in Fig. \ref{bc}. As happens for periodic boundary conditions, for helical boundary conditions, the states are determined by the configuration of two sets of $L$ sites, where the second one is shifted by one site only to the right of the first set, as is also shown in Fig. \ref{bc}. Thus, while for periodic boundary conditions at each application of the transfer matrix $L$ new sites are added to the lattice, for in the helical case a single site is added. This has the advantage of leading to a sparser matrix, since $L-1$ sites of both configurations are coincident. In Fig. \ref{bc}, two successive site sets, used to define the transfer matrix elements for helical boundary conditions, are enclosed by dashed and dotted rectangles.

\subsection{Number of states}

The numbers of states for lattices of successive widths are related by simple recursion relations. Let us call $n_{\sigma_1,\sigma_L}^{(L)}$ the number of states in a strip of width $L$ and free boundary conditions, such that the first site is in the configuration $\sigma_1$ and the last site in the configuration $\sigma_L$. Since $\sigma_i=0,1,2$, we have a total of 9 of these numbers of configurations, where 6 are independent due to reflection symmetry. The total number of configurations will be $n_f^{(L)}=\sum_{\sigma_1,\sigma_L} n_{\sigma_1,\sigma_L}^{(L)}$, for free boundary conditions. For periodic boundary conditions, the occupancy of the sites $1$ and $L$ should obey the excluded volume constraint, so that the number of states in this case will be given by $n_p^{(L)}=\sum_{\sigma_1,\sigma_L}^\prime n_{\sigma_1,\sigma_L}^{(L)}$, where the prime restricts the sum to terms such that $\sigma_1+\sigma_L \leq 2$. The numbers of states obey a set of nine linear recursion relations:
\begin{subequations}
\begin{eqnarray}
n_{\sigma_1,0}^{(L+1)}&=&n_{\sigma_1,0}^{(L)}+n_{\sigma_1,1}^{(L)}+n_{\sigma_1,2}^{(L)}, \\
n_{\sigma_1,1}^{(L+1)}&=&n_{\sigma_1,0}^{(L)}+n_{\sigma_1,1}^{(L)},\\
n_{\sigma_1,2}^{(L+1)}&=&n_{\sigma_1,0}^{(L)}.
\end{eqnarray}
\end{subequations}
Iterating these recursion relations with the starting values $n_{\sigma_1,\sigma_L}^{(1)}=\delta_{\sigma_1,\sigma_L}$, we may obtain the numbers of states for increasing values of the widths for both boundary conditions, which are shown in Tab. \ref{tab1}. The states for helical boundary conditions coincide with the ones for free boundary conditions. It is easy to reduce these recursion relations to a single one:
\begin{equation}
n_{\sigma_1,1}^{(L+1)}=2n_{\sigma_1,1}^{(L)}+n_{\sigma_1,1}^{(L-1)}-n_{\sigma_1,1}^{(L-2)},
\label{rr1}
\end{equation}
with the remaining numbers of states being given by $n_{\sigma_1,0}^{(L)}=n_{\sigma_1,1}^{(L+1)}-n_{\sigma_1,1}^{(L)}$, and
$n_{\sigma_1,2}^{(L)}=n_{\sigma_1,1}^{(L)}-n_{\sigma_1,1}^{(L-1)}$. The characteristic polynomial associated to the linear recursion relation (\ref{rr1}) will be:
$r^3-2r^2-r+1=0$, which has three real roots. Although of course the roots may be found analytically, we will not give the expressions here since they are rather long. The approximate numerical values are $r_1=2.2469797$, $r_2=-0.80193776$, and $r_3=0.55495811$. Therefore, in general, we have that:
\begin{equation}
 n_{\sigma_1,1}^{(L)}=A_{\sigma_1}r_1^L+B_{\sigma_1}r_2^L+C_{\sigma_1}r_3^L, 
\label{ns1}
\end{equation}
where the coefficients are determined by the initial conditions. 
To obtain the actual numbers of states for finite widths, it is more practical to iterate the recursion relations directly, but explicit expressions such at (\ref{ns1}) are useful to obtain the asymptotic number of states for $L \gg 1$, which is dominated by the leading root of the characteristic equation, $n^{(L)} \approx r_1^L \approx 2.2469797^L$. The coefficient of this asymptotic behavior will be different for free and periodic boundary conditions. This may be compared with the result for the model with large particles only, where we have, for periodic boundary conditions \cite{gb02}, 
that the numbers of states $n^{(L)}$ are given by a Fibonacci sequence starting with 1 and 3 $n_1{(L)}=F_{1,3}(L) \approx [(1+\sqrt{5})/2]^L  \approx 
1.6180 ^L$. As expected, the number of states increases much faster with the width when small particles are present.

\begin{table}[t]
\begin{tabular}{|c|c|c|}
\hline
$L$ & $n_f$ & $n_p$\\
\hline
   1    &           3   &         2 \\
   2    &           6   &         6 \\
   3    &          14   &        11 \\
   4    &          31   &        26 \\
   5    &          70   &        57 \\
   6    &         157   &       129 \\
   7    &         353   &       289 \\
   8    &         793   &       650 \\
   9    &        1782   &      1460 \\
  10    &        4004   &      3281 \\
  11    &        8997   &      7372 \\
  12    &       20216   &     16565 \\
  13    &       45425   &     37221 \\
  14    &      102069   &     83635 \\
  15    &      229347   &    187926 \\
  16    &      515338   &    422266 \\
  17    &     1157954   &    948823 \\
  18    &     2601899   &   2131986 \\
  19    &     5846414   &   4790529 \\
  20    &    13136773   &  10764221 \\

\hline
\end{tabular}
\caption{Numbers of states of the transfer matrix for strips of width $L$ for free ($n_f$) and periodic ($n_p$) boundary conditions.}
\label{tab1}
\end{table}

In order to reduce the amount of memory and computer time demanded, for periodic boundary conditions, we use a method proposed by Todo and Suzuki \cite{ts96} that consists in decomposing the matrix $T_{L}$ in $L+1$ sparse matrices, so that $T_{L} = \tilde{T}_{L}^{(L+1)} \cdot \tilde{T}_{L}^{(L)} \cdots \tilde{T}_{L}^{(2)} \cdot \tilde{T}_{L}^{(1)}$. Here, instead of adding a complete row in the strip, each matrix $\tilde{T}_{L}^{(i)}$, with $i=2,\ldots,L$ acts adding a new lattice site to the row. The matrix $\tilde{T}_{L}^{(1)}$ creates a site $i=1$ in a new row $h$, just above the site $i=2$ from the row $h-1$, accounting for the NN exclusion between particles in these sites. Then, $\tilde{T}_{L}^{(2)}$ adds the site $i=2$ in the row $h$ (above the site $[3,h-1]$), imposing the exclusion with particles in the site below it and the site $i=1$ in the row $h$. The other matrices up to $\tilde{T}_{L}^{(L)}$ work in the same way. Finally, the matrix $\tilde{T}_{L}^{(L+1)}$ imposes the NN exclusion between particles in the sites $L$ and $1$, and translate the labels $i \rightarrow i-1$, completing the formation of the row $h$. More details may be found in \cite{gb02}, where this method was applied to the case of large particles only. As an example of the power of this method, the number of non-vanishing terms in the symmetric matrix for $L=14$ is $M_{s}=31895812$, whereas the sum of such terms in the fifteen sparse matrices is $M_{sp}=6706321$, the ratio $r=M_{s}/M_{sp}$ is $r \approx 4.7$. For $L=16$ this ratio is $r \approx 15.6$ and for $L=18$ it is $r\approx 56.8$. This method allows us to work with strips of widths up to $L=18$ in computers with 16GB of RAM. The numerical method we used to find the two dominant eigenvalues of the transfer matrix was a variant of the power method \cite{w65}

After performing some numerical calculations for periodic and helical boundary conditions, we realized that the first, combined with the factorization of the transfer matrix, are more efficient, allowing us to handle strips of larger widths with the same computational effort. We will, therefore, present results for periodic boundary conditions only. Notice that to perform the factorization of the transfer matrix, no block-diagonalization using the symmetries is done.

\section{Results for periodic boundary conditions}
\label{resperiodic}

\subsection{Phenomenological renormalization}

\begin{figure}[!t]
\begin{center}
\includegraphics[width=8.5cm]{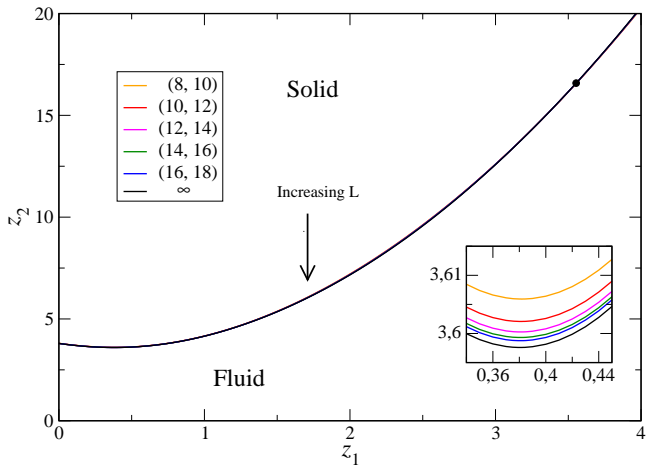}
\caption{(Color on line) Finite-size and extrapolated transition lines, calculated from the fixed point of the phenomenological renormalization transformation, considering the pair of strip sizes $L,L'=L+2$. In the main plot, the differences between the estimates are almost not visible. In the inset the region close to the minimum is amplified, so that the finite-size effects are seen. For a given value of $z_1$, as the width of the strip increases the estimates for the critical value of $z_2$ decrease. The estimated tricritical point, discussed below, is represented by the black circle.}  
\label{fig1}
\end{center}
\end{figure}

The inverse correlation length of the model for a strip of width $L$, is given by:
\begin{equation}
 \xi_{L}^{-1} = \ln \left( \frac{\lambda_{1}}{\lambda_{2}}\right),
\label{xi}
\end{equation}
where $\lambda_{i}$ is the $i^{th}$ largest eigenvalue of the corresponding transfer matrix. From previous results \cite{p84,tj11}, we may expect to find continuous and discontinuous transition lines between ordered and disordered phases. Phenomenological renormalization \cite{Ni76} states that the estimates for these transitions can be obtained from the fixed point of the recursion relation:
\begin{equation}
 \dfrac{\xi_{L}}{L} = \dfrac{\xi_{L'}}{L'}.
\label{pr}
\end{equation}
As discussed above, due to the symmetry of the crystalline phase, only even strip sizes must be considered, so that $L'=L+2$ is used. The transition lines are shown in Fig. \ref{fig1}, where a very small dependence on the size $L$ is observed. For a fixed value of $z_1$, we may expect this dependence to be:
\begin{equation}
 z_{2}(L) = z_{2c} + a_{1} L^{-x_1} + a_{2} L^{-x_2} + \ldots
\label{eq:correct}
\end{equation}
where $z_{2c}$ and the amplitudes $a_i$ should be functions of $z_1$, while the exponents are expected to be constant along the critical line. For $z_{1} = 0$, Guo and Bl\" ote \cite{gb02} found that $x_{i}=2+i$ and, in fact, for small $z_1$, we have verified that $x_1\approx 3$ and $x_2 \approx 4$. Then, assuming that these exponents hold for the whole transition curve, we obtain the values of $z_{2c}$ from Eq. (\ref{eq:correct}). The resulting transition line (TL) is also shown in Fig. \ref{fig1}, but no clear difference is observed from the ones for a given $L$ at this scale, due to the small finite-size corrections. Indeed, the maximal difference between the extrapolated TL and the line for the pair $(16,18)$ is $\approx 0.1$\%, which gives us an idea of the error bars in these estimates. In accordance with the Bethe lattice solution of the model \cite{tj11}, the critical line initially has a negative slope that approaches $-1$ when $z_1$ approaches zero. Increasing $z_1$, this curve reaches a minimum, located at $z_1 \approx 0.3805$ and $z_{2} \approx 3.5976$ - larger than the one found in the Bethe lattice ($z_1 \approx 0.26$ and $z_{2} \approx 1.55$) - and then starts to increase. Finally, we expect the curve to reach the tricritical point, beyond which the transition becomes discontinuous. The estimated tricritical point is also shown in the figure, and the methods we used to obtain this estimate will be discussed below.

\begin{table}[!t]
\begin{center}
\begin{tabular}{ccccc}

\hline\hline
$L$ & &  $z_{1}$  & &   $z_{2}$    \\
\hline
 6  & &  2.879392  & &  11.838332  \\
 8  & &  3.055064  & &  12.967494  \\
10  & &  3.194821  & &  13.918518  \\
12  & &  3.286902  & &  14.570770  \\
14  & &  3.348568  & &  15.019111  \\
16  & &  3.391375  & &  15.335858  \\
$\vdots$  & &  $\vdots$  & &  $\vdots$  \\
$\infty$  & &  3.559(2)  & &  16.60(4)  \\
\hline\hline
\end{tabular}
\caption{Estimates for the location of the tricritical point, using the condition $\dfrac{\xi_{L-2}}{L-2} = \dfrac{\xi_{L}}{L} = \dfrac{\xi_{L+2}}{L+2}$.}
\label{tab2}
\end{center}
\end{table}

\begin{figure}[!t]
\begin{center}
\includegraphics[width=8.5cm]{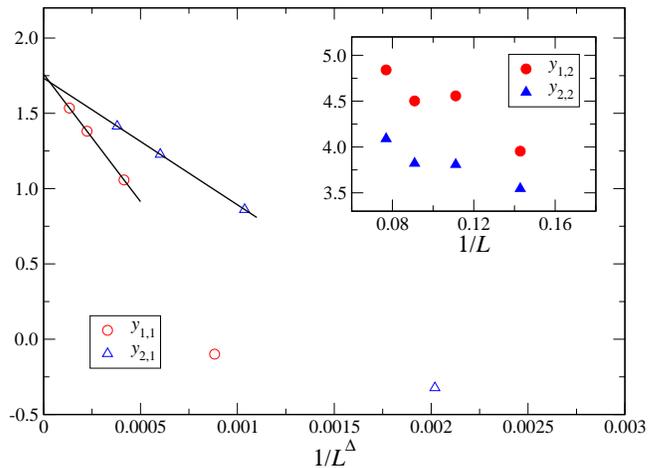}
\caption{(Color on line) Exponents of finite-size corrections (Eq. (\ref{eq:correctTC})) for the tricritical point activities estimated from phenomenological renormalization. In the main plot $y_{1,1}$ and $y_{2,1}$ are extrapolated with $\Delta=3.38$ and $\Delta=2.98$, respectively. The behaviors of $y_{1,2}$ and $y_{2,2}$ are shown in the inset.}  
\label{fig2}
\end{center}
\end{figure}

The estimates for the location of the tricritical (TC) point using results for finite-size strips may be obtained from the condition $\dfrac{\xi_{L-2}}{L-2} = \dfrac{\xi_{L}}{L} = \dfrac{\xi_{L+2}}{L+2}$ \cite{tcpc}. The values of $z_{1}^{TC}$ and $z_{2}^{TC}$ found for different $L$'s are show in Tab. \ref{tab2}. Assuming that
\begin{equation}
 z_{i}^{TC}(L) = z_{i}^{TC} + b_{i,1} L^{-y_{i,1}} + b_{i,2} L^{-y_{i,2}} + \ldots,
\label{eq:correctTC}
\end{equation}
where $i=1,2$, we may find the exponents $y_{i,1}$ considering that $b_{i,j}=0$ for $j>1$ and performing a three-point fit. Namely, we solve for $z_{i}^{TC}$, $b_{i,1}$ and $y_{i,1}$ in
\begin{equation}
 z_{i}^{TC}(L+l) = z_{i}^{TC} + b_{i,1} (L+l)^{-y_{i,1}},
\end{equation}
with $l=0,-2$ and $2$. The values of $y_{i,1}$ are not constant, but converge toward asymptotic values as $L$ increases, as shown in Fig. \ref{fig2}. Due to finite-size effects, for $L=8$, negative exponents are found, which do not make sense. Extrapolating the exponents for $L>8$, we find the same asymptotic value $y_{i,1}\approx 1.75$ for both $i=1$ and $i=2$ (see Fig. \ref{fig2}). Therefore, we will assume that $y_{i,1}=7/4$. Inserting this exponent in Eq. (\ref{eq:correctTC}) and assuming that $b_{i,j}=0$ for $j>2$, the value of $y_{i,2}$ can be obtained from a four-point fit (for the unknowns $z_{i}^{TC}$, $b_{i,1}$, $b_{i,2}$ and $y_{i,2}$). These exponents are shown in the inset of Fig. \ref{fig2}. Unfortunately, they do not present a monotonic behavior, possibly due to the small strip widths we could handle, so we cannot extrapolate them, but we may see that $y_{1,2} \gtrapprox 5$ and $y_{2,2} \gtrapprox 4$. Considering different exponents in the ranges $5 \leqslant y_{1,2} \leqslant 7$ and $4 \leqslant y_{2,2} \leqslant 6$, we obtain several estimates of the tricritical point, leading to $z_{1}^{TC}=3.557(3)$ and $z_{2}^{TC}=16.60(2)$. These values and their respective error bars are also calculated considering different $L$'s in the fit (not only the three largest strip sizes). We notice that these values of the activities at the tricritical point are much larger than the ones found on the Bethe lattice with coordination $q=4$ ($z_{1}^{TC}=1.16956$ and $z_{2}^{TC}=3.02938$) \cite{tj11}. It is, indeed, expected that mean-field approximations underestimate the critical activities.

Beyond the tricritical point the transition becomes discontinuous and, at first sight, we should not extrapolate the coexistence line in the same way we did for the critical line (below the TC point). Actually, from finite-size scaling, we expect exponential corrections at coexistence \cite{b84}, instead of the power law behavior (e. g., Eq. (\ref{eq:correct})) at criticality, namely
\begin{equation}
 z_{2}(L) = z_{2c} + a_1 e^{- b_1 L} + \ldots
\label{eq:correctexp}
\end{equation}
However, performing a three point fit considering this correction form, we have found a coexistence curve whose maximal difference from the one obtained as above (using Eq. (\ref{eq:correct})) is smaller than $0.02$\%. Thus, both approaches lead to very similar coexistence lines in the thermodynamic limit.

\subsection{Conformal anomaly}

\begin{figure}[!t]
\begin{center}
\includegraphics[width=8.5cm]{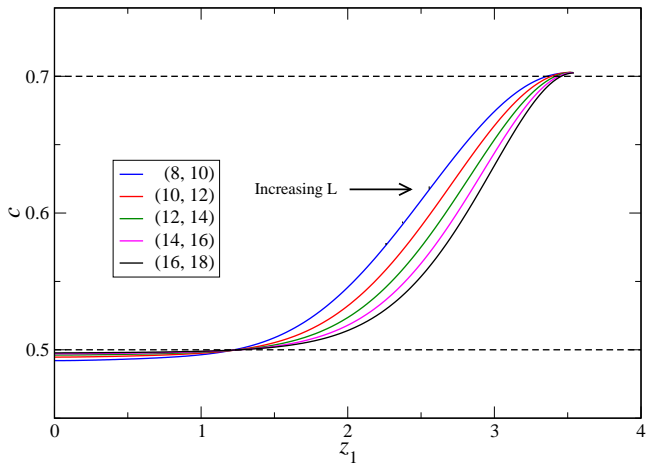}
\caption{(Color on line) Central charge $c$ as a function of $z_1$, calculated from $f(L')-f(L)$ along the finite-size critical line for the pairs $L,L'$ indicated. The dashed horizontal lines indicate the Ising critical (bottom) and tricritical (top) values.}
\label{fig3}
\end{center}
\end{figure}

The free energy per site of the system for a strip of width $L$ is given by
\begin{equation}
f(L) = \dfrac{1}{L} \ln \lambda_1.
\label{eq:freeenergy}
\end{equation}
Conformal invariance theory states that at criticality
\begin{equation}
f(L) \approx f_{\infty} + \dfrac{\pi c}{6 L^{2}},
\label{eq:energy}
\end{equation}
where $c$ is the central charge. Figure \ref{fig3} shows this quantity as a function of $z_1$ along the \textit{finite-size} critical lines and we can see a crossover from $c \approx 0.5$ at small $z_1$ to $c \approx 0.7$ near the TC point. Moreover, as the widths of the pair of strips become larger, the estimates for $c$ approach $c=1/2$ at low values of $z_1$ and the crossover is steeper. Therefore, in the thermodynamic limit we expect that $c=1/2$ for $z_1 < z_{1}^{TC}$ and $c=7/10$ at $z_1 = z_{1}^{TC}$, in good agreement with the critical and tricritical Ising universality classes, respectively, similar to what is found, for example, for the BEG model \cite{afks85}.

\begin{table}[!b]
\begin{center}
\begin{tabular}{ccccccc}
\hline\hline
$(L, L')$ & &  $z_{1}$  & &   $z_{2}$  & &   $c_{max}$ \\
\hline
$(6, 8)$  & &  3.461580  & &  15.844423  & &  0.700552  \\
$(8, 10)$  & &  3.503228  & &  16.172684  & &  0.702456  \\
$(10, 12)$  & &  3.522518  & &  16.327104  & &  0.702901  \\
$(12, 14)$  & &  3.532852  & &  16.410446  & &  0.702867  \\
$(14, 16)$  & &  3.538962  & &  16.459910  & &  0.702677  \\
$(16, 18)$  & &  3.542843  & &  16.491393  & &  0.702447  \\
$\vdots$  & &  $\vdots$  & &  $\vdots$  & &   \\
$\infty$  & &  3.5541(3)  & &  16.5827(3)  & &    \\
\hline\hline
\end{tabular}
\caption{Tricritical point estimates from the maxima of the central charge curves (Fig. \ref{fig3}) and the maximum values of the central charge $c_{max}$.}
\label{tab3}
\end{center}
\end{table}

The central charge curve, for a given pair $(L,L')$, has a maximum close to $c=7/10$ that should be located exactly at the tricritical point when $L \rightarrow \infty$. Thus, we may use the condition $\dfrac{\partial c}{\partial z_1} = 0$ as an alternative estimate of this point. The values obtained in this way are shown in Tab. \ref{tab3}. Performing extrapolations similarly to what was done in the previous subsection, we find $y_{i,1} \approx 2.37$. Moreover, exponents $y_{1,2}\gtrapprox 4.5$ and $y_{2,2}\approx 9.9$ are found. From a three-point extrapolation considering these exponents, we obtain $z_{1}^{TC}=3.5541(3)$ and $z_{2}^{TC}=16.5827(3)$, which agree with the estimates from phenomenological renormalization shown above, but are more accurate. These values will hereafter be used as the location of the TC point. 

Table \ref{tab3} also shows the central charge at the maximum as a function of $L$, where a non-monotonic behavior can be observed, preventing us to extrapolate this quantity in a reliable way. Nevertheless, the values in Tab. \ref{tab3} are very close to $c=7/10$ and they are approaching this limit for the larger widths.

\subsection{Densities of particles}

\begin{figure}[t]
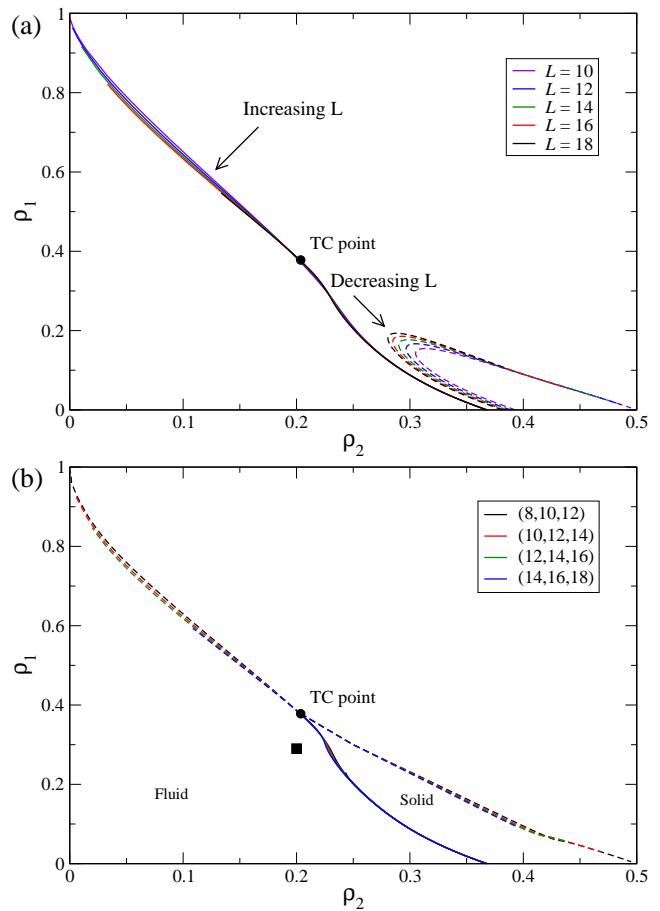

\begin{center}
\includegraphics[width=8.5cm]{Fig4a-rev.eps}
\includegraphics[width=8.5cm]{Fig4b-rev.eps}
\caption{(Color on line) (a) Transition (critical and coexistence) lines calculated for strips of size $L$ in the densities' space, for the fluid (full) and solid (dashed lines) phases . (b) Extrapolated critical (full) and coexistence (dashed) lines considering three point fits for the sizes $(L-2,L,L+2)$ and the exponents indicated in the text. The square is placed at the localization of the tricritical point estimated by Poland \cite{p84}.}
\label{fig4}
\end{center}
\end{figure}

The densities of particles for the fluid ($F$) and solid ($S$) phases can be obtained from 
\begin{equation}
 \rho^{F}_{i}(L) = z_{i} \frac{\partial f^{F}(L)}{\partial z_{i}}, \quad \text{and} \quad \rho^{S}_{i}(L) = z_{i} \frac{\partial f^{S}(L)}{\partial z_{i}},
\end{equation}
with $i=1,2$, and $f^{F}(L)$ and $f^{S}(L)$ being the free energies of the system in fluid and solid phases, respectively. We calculate these densities along the \textit{finite-size} transition lines (TL) in the following way: along the TL obtained for the pair $(L-2,L)$, we determine the densities for the strip width $L$. The resulting densities curves are displayed in Fig. \ref{fig4}a. For small $z_1$ (and $\rho_1$ - i. e., in the critical line) a very small finite-size dependence is observed in the curves for the $F$ phase, while for large $\rho_1$ (i. e, in the coexistence line) some $L$-dependence appears. Assuming finite-size corrections in the form
\begin{equation}
 \rho_{i}(L) = \rho_{i} + a_{i,1} L^{-v_{i,1}} + a_{i,2} L^{-v_{i,2}} + \ldots,
\label{eq:correctDens}
\end{equation}
from three and four point fits, we obtain the exponents $v_{i,1}\approx 2.8$ and $v_{i,2}\approx 4$. Then, using these exponents to extrapolate the transition line for the fluid phase, we find the extrapolated curves shown in Fig. \ref{fig4}b. It is worth noticing that as $z_1$ and, consequently, $\rho_1$ goes to zero (on the critical line) the density of large particles approaches the value $\rho_2=0.367742\ldots$ estimated by Guo and Bl\"ote \cite{gb02} for $\rho_1=0$.

Along the density curves for the solid phase, we may observe a re-entrant behavior, with the region below and above the turning point related to the critical and coexistence lines, respectively. At odds with the fluid phase, in the coexistence region (large $\rho_2$) negligible finite-size effects are observed, while strong corrections appears for small $\rho_2$ (in the critical curve). In this last region, below the turning point, if we assume that corrections are given by Eq. (\ref{eq:correctDens}), we find the exponents $v_{i,1}\approx 1$ and $v_{i,2}\approx 3$, which lead to extrapolated critical curves in very nice agreement with the ones for the fluid phase. This indicates that below the TC point we have in fact a critical situation where the densities of the two phases must be equal. Moreover, this confirms that the correction form assumed (Eq. (\ref{eq:correctDens})) with the exponents associated is in fact appropriate.

Although the Eq. (\ref{eq:correctDens}) still works on the coexistence line for the solid phase (large $\rho_2$), the exponents change to $v_{i,1}\approx 0.3$ and $v_{i,2}\approx 1.3$. The extrapolated coexistence curves with these exponents are also shown in Fig. \ref{fig4}b.

The densities of particles for strips of size $L$ calculated at the \textit{extrapolated} tricritical point  are shown in Tab. \ref{tab4}. It is interesting that, although the error bars are increasing, the values of the densities do not change with the strip size. This allows us to conclude that $\rho_1^{TC}=0.378(1)$ and $\rho_{2}^{TC}=0.2037(8)$ without extrapolations. As expected, these values are larger than the ones found in the Bethe lattice solution for coordination $q=4$ ($\rho_1 \approx 0.2985$ and $\rho_2 \approx 0.1117$).

\begin{table}[t]
\begin{center}
\begin{tabular}{ccccc}
\hline\hline
$L$ & & $\rho_{1}$  & & $\rho_{2}$ \\
\hline
10  & &  0.3780(5)  & &  0.2038(3)   \\
12  & &  0.3780(7)  & &  0.2037(4)   \\
14  & &  0.3780(9)  & &  0.2037(5)   \\
16  & &  0.378(1)   & &  0.2037(7)   \\
18  & &  0.378(1)   & &  0.2037(8)   \\
$\vdots$  &  &  $\vdots$   \\
$\infty$  &  &  0.378(1)  & &  0.2037(8)  \\
\hline\hline
\end{tabular}
\caption{Densities of small $\rho_1$ and large $\rho_2$ particles calculated at the tricritical point.}
\label{tab4}
\end{center}
\end{table}

\begin{figure}[t]
\begin{center}
\includegraphics[width=8.5cm]{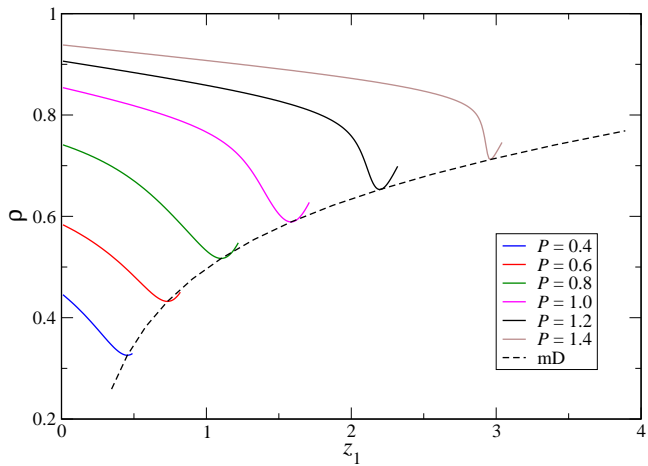}
\caption{(Color on line) Total density of particles $\rho = \rho_1 + 2 \rho_2$, calculated on a strip of size $L=18$, against $z_1$ for several values of pressure $P$. From the left to the right, curves correspond to increased pressures. The minima in the densities curves define the mD line (dashed).}
\label{fig5}
\end{center}
\end{figure}

In accordance with the solution on the Bethe lattice, the total density of particles for a fixed pressure - we may identify the reduced grand-canonical free energy $f_{\infty}$ in Eq. (\ref{eq:energy}) with the reduced  pressure - as a function of the fugacity $z_1$ or of the density $\rho_1$ of small particles has a non-monotonic behavior, it displays a minimum in the fluid phase, as shown in Fig. \ref{fig5}. These curves may be obtained fixing the value of the dominant eigenvalue of the transfer matrix (this corresponds to a fixed value of the pressure), so that $z_2$ and $\rho$ may be found as functions of $z_1$. Since at larger widths the finite size effects in these calculations are rather small, we have done them for strips of width $L=18$ without extrapolations. The locations of these minima originate a curve similar to the temperature of maximal density (TMD) curve delimiting the density anomaly in water-like fluids \cite{c06}, so we will call these minima as mD (minimal density) points. Notice that in the case of water, the curves which exhibit a maximum are the isobars of the density as a function of the temperature. Here we also have isobars, but the density is expressed as a function of an activity, which is also a field-like variable in the thermodynamic sense, or of the conjugated density, which is a monotonic function of the activity. The mD curves seem to start at $z_1=1/3$ as $z_2\rightarrow 0$ and then they are increasing functions of $z_1$. Although it is very difficult to calculate the minima close to the tricritical point, it seems that the curves end exactly at this point. For small $z_1$, the mD's obtained for different strip sizes have negligible corrections, but close to the tricritical point they present an appreciable $L$-dependence. Thus, we estimated the mD curve in the thermodynamic limit using a three-point fit.

\subsection{Phase diagrams}

\begin{figure}[t]
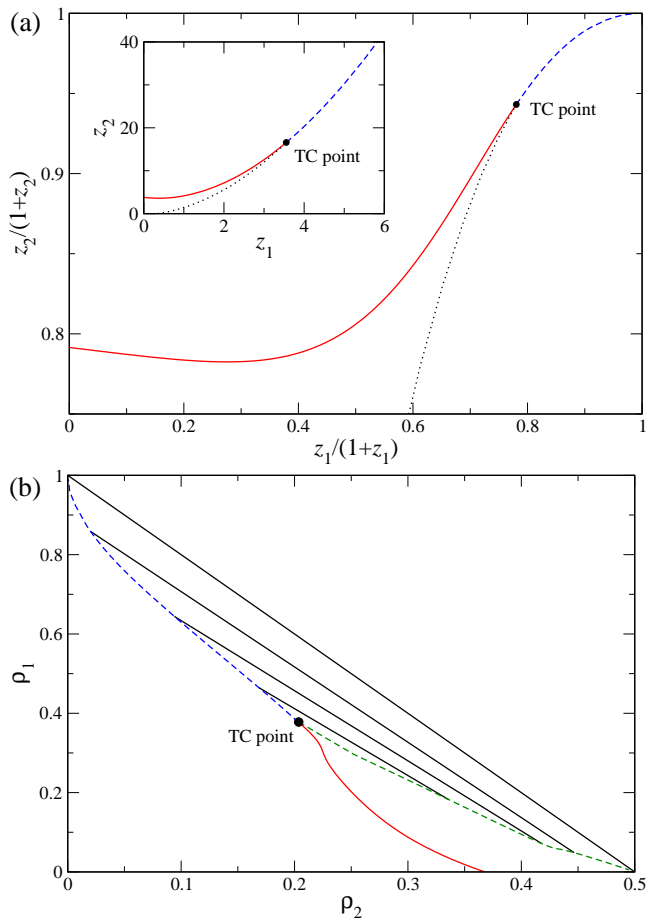

\begin{center}
\includegraphics[width=8.5cm]{Fig6a-rev.eps}
\includegraphics[width=8.5cm]{Fig6b-rev.eps}
\caption{(Color on line) (a) Phase diagram in the variables $z_i/(1+z_i)$ (main plot) and $z_i$ (inset), with $i=1,2$. Critical, coexistence and mD lines are indicated by continuous (red), dashed (blue) and dotted (black) lines, respectively. (b) Phase diagram in the densities $\rho_1$ and $\rho_2$ space. The densities of fluid (blue) and solid (green) phases at the same point of the coexistence line are connected by tie lines.} 
\label{fig6}
\end{center}
\end{figure}

Figure \ref{fig6} summarizes our results, showing the extrapolated critical line, coexistence line and mD curve in the reduced activity fraction variables $z_i/(1+z_i)$ in the main plot and activities $z_1$, $z_2$ in the inset. The phase diagram in the density variables is shown in Fig. \ref{fig6}b. Although these phase diagrams are qualitatively similar to the ones we obtained previously in the Bethe lattice solution of the model \cite{tj11}, as expected the transition lines are shifted to larger activities (or densities) on the square lattice. Also, we notice that the lines of the densities of coexisting phases meet at an angle in the Bethe lattice solution, while in our results for the square lattice they meet tangentially. This is expected, since the classical value of the tricritical exponent associated to the behavior of the order parameter $\rho$ is $\beta_2=1$ and the estimates for this exponent in two dimensions, one below the upper tricritical dimension $d=3$, is smaller than 1 \cite{ls84}.

\section{Conclusion}
\label{conc}

In this paper we study a model of a mixed lattice gas with two kinds of particles. Small particles exclude only the site they occupy, while large particles exclude, besides their site, its four (square lattice) first neighbors also. The model with only large particles is well studied in the literature \cite{b60,b80,gb02} and displays a continuous transition between a fluid phase where sites of the lattice are occupied at random to a solid phase where one of the two sub-lattices is preferentially occupied by the particles. As the small particles are introduced, a line of continuous transitions is found, which ends at a tricritical point, beyond which the transition is discontinuous. Using transfer matrix and finite-size scaling techniques, we estimate the thermodynamical behavior of the model extrapolating data of its solutions on strips of finite widths to the two-dimensional limit. 

The phase diagrams we found both in activity variables and densities spaces are qualitatively similar to the ones we found before in the Bethe lattice solution \cite{tj11}, but some characteristics close to the tricritical point are distinct, since at two dimensions non-classical tricritical exponents are expected, leading to changes as compared to the mean-field behavior, as was discussed above. It is interesting to compare our results with the ones obtained by Poland using high density series expansions \cite{p84}. Our estimate for the density of small particles at the TC point ($\rho_1^{TC}=0.378(1)$) is higher than the value obtained by Poland ($\rho_1^{TC}=0.29 \pm 0.02$), but our estimate for the density of large particles at this point ($\rho_2^{TC}=0.2037(8)$) is consistent with the one he found ($\rho_2^{TC}=0.20 \pm 0.01$). We notice that we obtained estimates for the localization of the tricritical point with different methods, which lead to consistent results. In particular, the estimate by Poland is not close to our results for the critical line in the density variables, it is below all curves shown in Fig. \ref{fig4}, where it is represented by the black square. Also, the critical line in the phase diagram in the activity variables, Fig. \ref{fig6}a, has a negative slope at small values of $z_1$, which becomes positive after a minimum. Thus, the same re-entrant behavior found in the Bethe lattice solution of the model \cite{tj11} is also found in our results for the square lattice. This shows that, when their density is low, the small particles facilitate the ordering of the large ones - namely, they acts as an effective entropic attractive force among the large particles - regardless the structure of the underlying lattice. It may be mentioned that in a study of the isotropic-nematic transition for polydisperse rods, the critical curve also has a non-monotonic behavior in the space of two fugacities, although the mean length of the rods changes monotonically along the curve \cite{sr15}.

Our estimates for the central charge are consistent with the Ising universality class on the whole critical line, in accordance with the very precise results obtained with similar techniques by Guo and Bl\"{o}te for the model at the particular point of the critical line with large particles only \cite{gb02} and with results of a similar model which may be mapped, in a particular case, on the Ising model \cite{fl92}. The estimates for the central charge at the tricritical point are close to the value of the tricritical Ising (BEG) universality class.

\section*{Acknowledgments}

We thank Ronald Dickman for having called our attention to this model and for discussion, and R. Rajesh for a critical reading of the manuscript. Partial funding from the Brazilian agencies CNPq and FAPEMIG are acknowledged.


\begin{thebibliography}{99}
\bibitem{hd86} J. P. Hansen and I. R. McDonald {\em Theory of Simple Liquids}, Academic Press (1986).

\bibitem{mrrtt53} N. Metropolis et al, J. Chem Phys. {\bf 21}, 1087 (1953).

\bibitem{b60} D. M. Burley, Proc. Phys. Soc. {\bf 75}, 262 (1960) and {\bf 77}, 451 (1961); D. A.  Gaunt and M. E. Fisher, J. Chem. Phys. {\bf 43}, 2840 (1960); L. K.  Runnels, Phys. Rev. Lett. {\bf 15}, 581 (1965);F. H.  Ree and D. A. Chesnut, J. Chem. Phys. {\bf 45}, 3983 (1967).

\bibitem{b80} R. J. Baxter, I. G. Enting, and K. S. Tsang, J. Stat. Phys. {\bf 22}, 465 (1980).

\bibitem{gb02} W. Guo and H. W. J. Bl\"ote, Phys. Rev E {\bf 66}, 046140 (2002).

\bibitem{fal07} H. C. M. Fernandes, J. J. Arenzon, and Y. Levin, J. Chem. Phys. {\bf 126}, 114508 (2007), and references therein.

\bibitem{rajesh14} T. Nath and R. Rajesh, Phys. Rev. E {\bf 90}, 012120 (2014), and references therein.

\bibitem{hb96} J. R. Heringa and H. W. J. Bl\"{o}te, Physica A {\bf 232}, 369 (1996).

\bibitem{zd08} W. Zhang and Y. Deng, Phys. Rev. E {\bf 78}, 031103 (2008).

\bibitem{p84} D. Poland, J. Chem. Phys. {\bf 80}, 2767 (1984).

\bibitem{fl92} D. Frenkel and A. A. Louis, Phys. Rev. Lett. {\bf 68}, 3363 (1992).

\bibitem{lt94} J. C. Lin and P. L. Taylor, Phys. Rev. Lett. {\bf 73}, 2863 (1994). See also J. M. Romero-Enrique, I. Rodr\'{\i}guez-Ponce, L. F. Rull, and U. M. B. Marconi, Phys. Rev. Lett. {\bf 79}, 3543 (1997).

\bibitem{wa80} J. C. Wheeler and G. R. Anderson, J. Chem. Phys. {\bf 73}, 5778 (1980).

\bibitem{tj11} T. J. Oliveira and J. F. Stilck, J. Chem. Phys. {\bf 135}, 184502 (2011).

\bibitem{s10} J. N. da Silva, E. Salcedo, A. B. de Oliveira, and M. C. Barbosa, J. Chem. Phys. {\bf 133}, 244506 (2010).

\bibitem{ts96} S. Todo and M. Suzuki, Int. J. Mod. Phys. C {\bf 7}, 811 (1996).

\bibitem{w65}J. H. Wilkinson, {\em The algebraic eigenvalue problem}, Oxford University Press (1965).

\bibitem{Ni76} M. P. Nightingale, Physica A {\bf 83}, 561 (1976).

\bibitem{tcpc} B. Derrida and H. J. Hermann, J. Phys (Paris) {\bf 44}, 1365 (1983).

\bibitem{b84}M. N. Barber in {\em Phase Transitions and Critical Phenomena}, vol. 8, ed. by C. Domb and J. L. Lebowitz, Academic Press (1984).

\bibitem{afks85} M. L\"{a}ssig, G. Mussardo, and J. L. Cardy, Nucl. Phys. {\bf B348}, 591 (1990). See also F. C. Alcaraz, J. R. Drugowich de Fel\'{\i}cio, R K\"{o}berle, and J. F. Stilck, Phys. Rev. B {\bf 32}, 7469 (1985) for a transfer matrix study of the tricritical point in the BEG model.

\bibitem{c06} M. Chaplin, Sixty-three anomalies of water,
http://www.lsbu.ac.uk/water/anmlies.html (2006).

\bibitem{ls84} I. D. Lawrie and S. Sarbach in {\em Phase Transitions and Critical Phenomena}, vol. 9, ed. by C. Domb and J. L. Lebowitz, Academic Press (1984).

\bibitem{sr15} J. F. Stilck and R. Rajesh, Phys. Rev. E {\bf 91}, 012106 (2015).

\end{thebibliography}
\end{document}